\documentclass[10pt]{iopart}
\usepackage{graphics}
\newcommand{\tu}{\tilde{u}}
\begin{document}
\title{Fluctuating-friction molecular motors}
\author{L. Marrucci\dag, D. Paparo\dag and M. Kreuzer\ddag}
\address{\dag\ INFM and Universit\`{a} ``Federico II'', Dip.\ di Scienze
Fisiche, Monte S.Angelo, v.\ Cintia, 80126 Napoli, Italy}
\address{\ddag\ Darmstadt University of Technology, Institute of Applied Physics,
Hochschulstrasse 6, 64289 Darmstadt, Germany}
\begin{abstract}
We show that the correlated stochastic fluctuation of the friction
coefficient can give rise to long-range directional motion of a
particle undergoing Brownian random walk in a constant periodic energy
potential landscape. The occurrence of this motion requires the
presence of two additional independent bodies interacting with the
particle via friction and via the energy potential, respectively,
which can move relative to each other. Such three-body system
generalizes the classical Brownian ratchet mechanism, which requires
only two interacting bodies. In particular, we describe a simple
two-level model of fluctuating-friction molecular motor that can be
solved analytically. In our previous work [M.K., L.M and D.P. 2000
{\it J. Nonlinear Opt.\ Phys.\ Mater.} {\bf 9} 157] this model has
been first applied to understanding the fundamental mechanism of the
photoinduced reorientation of dye-doped liquid crystals. Applications
of the same idea to other fields such as molecular biology and
nanotechnology can however be envisioned. As an example, in this paper
we work out a model of the actomyosin system based on the
fluctuating-friction mechanism.
\end{abstract}
\maketitle
\section{Introduction}
Driven by the modern tools of molecular biology and by the opening
perspectives of nanotechnology, there is currently a strong interest
into understanding all the mechanisms by which systems at a molecular
scale can efficiently convert chemical or light energy into mechanical
energy. Such systems are often called ``molecular motors'', and are at
the root of biological processes such as muscle contraction, cell
motility, and several intracellular transport processes
\cite{huxley57,hill74,svoboda93,finer94,yin95,noji97}. Contrary to
ordinary engines, these molecular motors are conceived to work at a
single temperature. Moreover they essentially exploit Brownian motion,
converting its random behaviour into an ordered directional motion by
means of some physical mechanism related with energy dissipation. An
example of such a mechanism is the so-called ``Brownian ratchet
effect'' \cite{magnasco93,astumian97,julicher97}. Artificial Brownian
ratchets at microscopic scales have been proposed for a variety of
technological applications, and in some cases they have been
experimentally demonstrated
\cite{drexler,rousselet94,faucheux95,zapata96,lee99}. Very recently,
the first steps toward engineering truly molecular motors have been
taken \cite{kelly99,koumura99}.

Within this field, a specific line of investigation is aimed at developing
physical models that capture the essential features of these molecular
systems but that are simple enough to be easily understood and studied
\cite{julicher97}. Many models are focused on a very simple ideal system: a
point-like particle undergoing overdamped Brownian motion under the effect
of molecular friction and of a periodic energy potential $U(x)$, where $x$
is the particle coordinate. At a given constant absolute temperature $T$,
friction and Brownian diffusion are characterized by a single quantity that
can be taken to be the friction coefficient $\eta$ or equivalently the
diffusion constant $D=kT/\eta$, where $k$ is Boltzmann constant. In this
case, it is well known that even if $U(x)$ is asymmetric for
$x\rightarrow-x$, as for example in the case of a saw-tooth potential, no
directional long-range motion can be induced unless the system is driven
out of thermodynamic equilibrium by some mechanism \cite{feynmann}.

One of the simplest proposed mechanisms to induce directional motion is a
random switching of the system between two internal states characterized by
different potentials, say $U_1(x)$ and $U_2(x)$ \cite{prost94} (see also
\cite{oester95,oester95b} for a detailed two-state model of kinesin). In
other words, the particle is subject to a fluctuating potential, where the
fluctuations are described as sudden switches from $U_1(x)$ to $U_2(x)$ and
backwards. These switches are not completely random, but obey to stochastic
laws in which detailed balance and hence thermodynamic equilibrium is
broken. For example, one must assign the probability per unit time
$I_{ij}(x)$ of having a nonthermal forced transition from state $i$ to
state $j$, associated with some input of free energy. Moreover, there is
the probability per unit time of having spontaneous transitions associated
with thermal equilibrium and therefore obeying to detailed balance. One may
prove for this system that the correlated energy potential fluctuations
associated with the random transitions may result into a nonzero average
force $F$ acting on the particle and therefore into its drift at a constant
average velocity $v=F/\eta$ \cite{prost94}. This long-range motion is the
mechanical output of the motor, whereas the (chemical or optical) free
energy input arises from the nonthermal state transitions. The direction of
this motion is dictated by the asymmetry of one or both the potentials
$U_i(x)$. However, a directional motion can be induced also by asymmetric
transition rates \cite{julicher97}.

The possibility of a difference in the friction coefficient (or diffusion
constant) experienced by the particle in the two internal states, i.e.,
$\eta_1\neq\eta_2$ (or $D_1{\neq}D_2$), was considered in previous works
but only with a passive role. In particular, if taken alone, this
difference cannot give rise to a nonzero average force and velocity of the
motor. In other words a fluctuation of friction in the presence of a
constant force potential cannot power the motor.

In this paper we show that a stochastic fluctuation of friction
induced by asymmetric transition rates actually can give rise to a
nonzero average force $F$. Unexpectedly, however, this steady average
force does not generate any long-range motion of the particle, i.e.,
$v=0$. More precisely, it can be shown that the average friction force
experienced by the particle, owing to the fluctuations in the friction
coefficient, does not vanish even for $v=0$. Therefore, a balance
between the average nonzero force arising from the (constant)
potential $U(x)$ and the average friction force is established at
$v=0$. At first sight it seems therefore that there is no possibility
at all of obtaining motion and mechanical work out of Brownian motion
with fluctuating friction coefficient. We show in the following that
this is not completely true.

Let us consider explicitly the two ``external'' bodies that interact with
the particle and provide respectively the potential forces and the friction
forces acting on the particle. In the following, we call these two bodies
the ``sources'' of the potential and friction forces. Usually, in
discussing molecular motor models, these bodies are considered as large
systems that do not move. However, this picture is not always realistic.
For example, in biological actomyosin motors the actin (``thin'') filaments
(usually considered as the source of potential forces in the framework of
these models) are displaced relative to myosin (``thick'') filaments by the
action of many myosin motors. It is this displacement the actual output of
the motor. Therefore, we are led naturally to consider also the motion of
these ``large'' bodies, besides the particle itself, in analyzing the
mechanical output of these systems. The whole motor system must actually be
seen then as a three-body dynamical system.

Now, the bodies generating the opposite friction and potential forces
acting on the particle are also, by reaction, experiencing opposite
nonvanishing average forces. Therefore these two large bodies can be set in
motion relative to each other, unless they are blocked by some constraint
or they actually coincide. In other words, the motor particle undergoing
Brownian motion with a fluctuating friction coefficient can induce a
long-range relative displacement of the two bodies with which it is
interacting. In the process, the particle itself may also undergo a
directional long-range displacement.

This paper is organized as follows. In the next Section we describe a
simple model of a three body-system that converts chemical (or
optical) free energy into mechanical work by exploiting the Brownian
motion of a particle that experiences stochastic sudden fluctuations
of its friction coefficient between two values. In section III we
solve analytically the equations of this system for a specific choice
of the potential landscape and transition probabilities. We also
performed some numerical studies to analyze more general cases. In
section IV, in order to show how the idea of a fluctuating-friction
molecular motor can be useful in the context of biological systems, we
apply our model to the actomyosin system. Some concluding remarks are
given in section V.

\section{The model}
Consider three bodies denoted as A, B, and C, corresponding to the
motor active unit (the ``particle''), the potential source, and the
friction source, respectively (see figure 1). We start working in a
frame of reference $\cal{R}$ (we may consider it ``fixed'', for the
sake of simplicity) in which C is motionless while A and B move, their
motion being described by coordinates $x_A(t)$ and $x_B(t)$. The
motion of B is supposed to be deterministic, i.e., its Brownian
fluctuations are neglected, for example because it is a comparatively
larger body. For the time being, we assume that B moves at a constant
velocity $V_B$, i.e., $x_B=V_Bt$. The motion of the small particle A
is instead stochastic, due to Brownian fluctuations. Moreover, we
assume that it is overdamped, i.e., inertia of A can be neglected.
Therefore, to each internal state $i=1,2$ of the motor we may
associate an evolving probability density $f_i(x_A,t)$. The
interaction between A and B is described by a potential
$U_i(x_A,x_B)=U_i(x_A,V_Bt)$ that is time-dependent, because B is
moving. It is therefore convenient to switch to another inertial frame
of reference ${\cal{R}}_B$ that is comoving with B, thus introducing a
new relative coordinate $x=x_A-x_B$ in terms of which the potential
$U_i(x)$ is stationary.
\begin{figure}
\begin{center}
\includegraphics{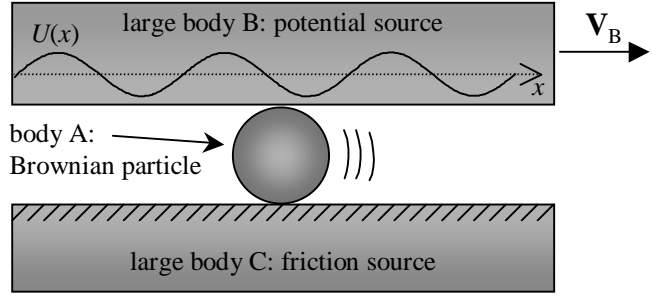}
\end{center}
\caption{Schematic drawing of our three body model of fluctuating-friction
molecular motor. The potential $U(x)$ characterizes the interaction between
body B and the particle A, where $x$ is their relative position. Body C
interacts with the particle via friction.}
\end{figure}

The stochastic dynamics of body A can then be described by the following
pair of coupled Smoluchowski equations:
\begin{equation}
\frac{\partial f_i}{\partial t}+\frac{\partial J_i}{\partial x} = W_i(x),
\label{maineq}
\end{equation}
where $J_i(x)$ is the probability current in the frame ${\cal{R}}_B$ and
$W_i(x)$ is the net rate of transitions to state $i$ at position $x$. $J_i$
is given by
\begin{equation}
J_i = -D_i \left( \frac{\partial f_i}{\partial x} +
\frac{f_i}{kT}\frac{\partial U_i}{\partial x}\right) - f_i V_B, \label{eqJ}
\end{equation}
where $D_i=kT/\eta_i$ is the diffusion constant in state $i$. The
first two terms in equation (\ref{eqJ}) are the standard diffusion and
drift currents, respectively. The last term appears because we are
working in a frame of reference, ${\cal{R}}_B$, that is moving with
respect to the source of friction (normally, overdamped Brownian
motion is studied in the privileged reference frame of the fluid
providing the friction). Formally, it can be derived starting from
equations (\ref{maineq}) written in the fixed frame $\cal{R}$, by
applying the Galilei coordinate transformation $x=x_A-V_Bt$ to the
$\partial{f_i}/\partial{t}$ term. Physically, it describes the drift
current induced by the drag force generated by the friction source
body C that -- in the frame ${\cal{R}}_B$ -- is moving at constant
speed $-V_B$. This term would arise, for example, when describing the
Brownian motion of a particle in a viscous fluid that is flowing at
constant speed $-V_B$.

For comparing the contributions of fluctuating friction and fluctuating
potential, we allow here for both of them to depend on the internal state.
For the transition rates we take
\begin{equation}
W_2(x)=-W_1(x)=I(x)f_1(x)-\frac{f_2(x)}{\tau},
\end{equation}
with $\tau$ being the lifetime of state 2 (state 1 is assumed to be stable,
while $1/\tau$ is the rate of spontaneous transitions $2\rightarrow1$). The
latter follows from assuming that state transitions are local, i.e., do not
involve a variation of $x$, and thermal-induced transitions from level 1 to
level 2 are negligible (because the level energy difference is much larger
than $kT$). Note moreover that we are assuming $I(x)$ to depend only on the
relative coordinate $x=x_A-x_B$.

Thus far we have made no hypothesis on the periodicity of $U(x)$ and
$I(x)$. Actually, this is not a strictly necessary condition in our model,
as we will discuss later. However, we assume in the following that the
functions $U_i(x)$ and $I(x)$ are periodic, owing to the extended periodic
structure of body B. Denoting then by $L$ the half-period in the variable
$x$, we may limit our solution to the interval $x\in[-L,L]$ by imposing the
periodic boundary conditions $f_i(-L)=f_i(L)$ and $J_i(-L)=J_i(L)$. The
distribution functions $f_i$ are normalized by the condition
$\sum_i\int_{-L}^Lf_idx=1$.

In the stationary regime, when $\partial f_i/\partial t=0$, the
condition $W_1=-W_2$ combined with equations (\ref{maineq}) implies
that the total current $J_t=J_1(x)+J_2(x)$ is independent of $x$. This
first-integral can be exploited to reduce equations (\ref{maineq}) and
(\ref{eqJ}) to a set of three first-order differential equations in
the unknown functions $f_1(x),f_2(x)$, and $J_2(x)$, with $J_t$
playing the role of an eigenvalue. Then, the three remaining periodic
boundary conditions together with the normalization condition uniquely
determine the unknown functions and the eigenvalue $J_t$. In this way,
the problem can be easily solved numerically for any functional shape
of $U_i(x)$ and $I(x)$ and for a given velocity $V_B$. For the case of
piecewise linear functions shown in figure 2, the problem can also be
solved analytically, as discussed in section III.
\begin{figure}
\begin{center}
\includegraphics{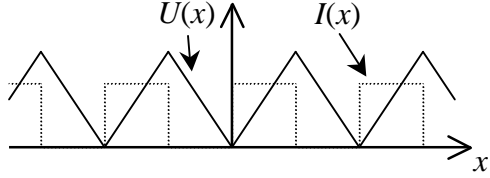}
\end{center}
\caption{A particular choice of the potential $U(x)$ and transition probability
$I(x)$ that allows a fully analytical solution of our model. Notice that,
owing to the relative phase of the two functions, the $x\rightarrow-x$
symmetry is broken. This symmetry breaking determines the direction of the
long range motion.}
\end{figure}

Once solved the stationary random-walk problem, we are interested in
computing the average force $F_{AB}$ that B exerts on A, given by
\begin{equation}
F_{AB} = - \sum_i\!\int_{-L}^L \frac{dU_i}{dx} f_i dx,
\end{equation}
and the average friction force $F_{AC}$ arising in the interaction with C,
given by
\begin{equation}
F_{AC} = -\sum_i\frac{kT}{D_i}\!\int_{-L}^L \!\left(J_i+f_iV_B\right) dx
\label{eqF1}
\end{equation}
(the latter is easily understood by considering that $J_i+f_iV_B$ is
the current in the fixed frame $\cal{R}$, i.e., the current relative
to body C). By exploiting equations (\ref{eqJ}), it is straightforward
to prove that $F_{AB}+F_{AC}=0$, reflecting the balance of all forces
acting on the element A of the motor. Due to this balance, the
particle A does not acquire any momentum (that it could not
accumulate, since its inertia is negligible). On the other hand, by
interacting with B and C, A mediates a continuous transfer of momentum
from the friction source body C to the potential source body B. This
transfer is expressed by the reaction force $F_{BA}=-F_{AB}=F_{AC}$
acting on B. In the following, for brevity, we denote the latter just
by $F=F_{BA}$. A convenient expression of $F$, obtained from equation
(\ref{eqF1}), is the following:
\begin{equation}
F(V_B)=-\frac{2kTL}{D_1}\left[J_t+\left(\frac{D_1}{D_2}-1\right)
\bar{J}_2\right]-\bar{\eta}V_B,
\label{eqF2}
\end{equation}
where
\begin{equation}
\bar{J}_2=\frac{1}{2L}\int_{-L}^LJ_2(x)dx
\end{equation}
is the average current in state 2 and
\begin{equation}
\bar{\eta}=kT\int_{-L}^L\left(\frac{f_1(x)}{D_1}+\frac{f_2(x)}{D_2}\right)dx
\end{equation}
is the average friction coefficient.

The average relative velocity $v$ of the active element A in the frame
${\cal{R}}_B$ can be evaluated as $v=2LJ_t$. If we set $V_B=0$ and there
are no potential fluctuations, i.e., $U_1(x)=U_2(x)=U(x)$, then $v=J_t=0$,
for any functional form of $U(x)$. To prove it, let us introduce the
function
\begin{equation}
g(x)=\left[D_1f_1(x)+D_2f_2(x)\right]e^{U(x)/kT}.
\end{equation}
The periodic boundary conditions and the periodicity of $U(x)$ imply
that $g(-L)=g(L)$. However, from equations (\ref{eqJ}) we find the
relation
\begin{equation}
\frac{dg}{dx}=-J_te^{U(x)/kT},
\end{equation}
which after an integration gives
\begin{equation}
J_t=\frac{g(-L)-g(L)}{\int_{-L}^Le^{U(x)/kT}dx}=0.
\end{equation}
This means that there can be no long range displacement of the
particle A with respect to the potential source for $V_B=0$, as
anticipated in the Introduction. From equation (\ref{eqF2}) one can
see, however, that the vanishing of $J_t$ does not imply the vanishing
of the force $F$ if $D_1{\neq}D_2$, as the average current $\bar{J}_2$
does not necessarily vanishes for $V_B=0$. In fact, as illustrated
graphically in figure 3, the system may achieve a stationary
nonequilibrium state in which there is a continuous cycling between
state 1, in which the particle in the average moves one way, and state
2, in which in the average the particle moves the opposite way by the
same amount. The net particle displacement vanishes but, owing to the
different friction coefficients in the two states, the average
friction force does not. The actual achievement of this stationary
state is proved in section III for the particular choice of $U(x)$ and
$I(x)$ shown in figure 2. In general, we found numerically that $F$
does not vanish if the nonthermal transition probability $I(x)$ is
described by a periodic function that is shifted with respect to the
potential $U(x)$, so that the resulting overall transition rates
$W_i(x)$ are asymmetric. Note that if one sets $V_B\neq0$, then
$J_t\neq0$, i.e., particle A will acquire a directional motion
relative to the other two bodies.
\begin{figure}
\begin{center}
\includegraphics{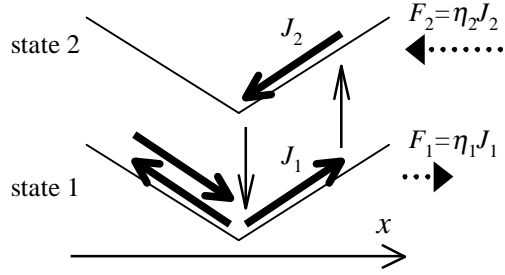}
\end{center}
\caption{Probability-flux loop due to state-transitions and probability
currents in the Brownian motion of body A, for the case of no
potential fluctuations ($U_1=U_2=U(x)$). The potential $U(x)$ and the
transition probability $I(x)$ are as in figure 2. Although the total
current vanishes, because $J_2=-J_1$, the contribution to the total
friction force $F=F_1+F_2=\eta_1J_1+\eta_2J_2$ is nonzero if
$\eta_1\neq\eta_2$.}
\end{figure}

Let us now come back to body B. Thus far we have simply assumed that B is
moving at constant speed $V_B$, and we have thus determined the average
force $F=F(V_B)$ that the motor develops on B itself. Now we can use this
information to write a self-consistent equation for the (deterministic)
motion of B at steady-state,
\begin{equation}
m_B\frac{dV_B}{dt}=F(V_B)-\eta_pV_B-F_{\mathrm{ext}}=0, \label{eqB}
\end{equation}
where $\eta_p$ is an additional friction coefficient associated with
the motion of body B, and $F_{\mathrm{ext}}$ is the external load. In
particular, setting $V_B=0$ in equation (\ref{eqB}) we obtain the
``stalling'' force $F_{\mathrm{ext}}=F(0)$. The maximum velocity
$V_{\mathrm{max}}$ is instead obtained by solving the force-balance
equation with $F_{\mathrm{ext}}=0$.

Note that we are assuming here that body B responds only to the average
force $F(V_B)$, and not to the instantaneous fluctuating force exchanged
with A. As we said, this assumption is justified for example if B is much
larger and therefore slower than A, so that it responds only to its
time-averaged motion and ignores its fast Brownian fluctuations. Another
case that justifies this approach occurs when B interacts simultaneously
with many replicas of A and therefore it responds only to their total force
(proportional to the average force of a single motor). However, it should
be noted also that this simplifying assumption of our model is not
essential to our main conclusions and that qualitatively similar results
are to be expected when Brownian fluctuations of B are taken into account.

One final comment is in order. At first sight, one could be led to
believe that by assuming from the start a nonzero velocity for body B
we have artificially introduced a bias in the random-walk of A, and
that it is only this bias that finally leads to the directional
motion. That this is not the case is proved by the fact that the
internal force $F(V_B)$ does not vanish for $V_B=0$. It is this force
that defines the long-range motion direction. And it is this internal
force that, if not counteracted by a stalling external load,
eventually sets body B in motion.

\section{Solutions}
Let us now describe some specific solutions of our model. It is
convenient to use dimensionless quantities, obtained from the
corresponding dimensional ones by using $L,L^2/D_2,kT$ as units of
length, time, and energy, respectively. We denote a dimensionless
quantity by a tilde sign placed above the corresponding dimensional
symbol. In particular $\tilde{I}(x)=I(x)L^2/D_2$, $\tilde{F}=FL/kT$,
$\tilde{D}_1=D_1/D_2$, and $\tilde{V}_B=V_BL/D_2$. For definiteness,
let us specialize to the case of triangular-wave potentials and
square-wave transition probability shown in figure 2, i.e.,
$\tilde{U}_i(\tilde{x})=[2\theta(\tilde{x})-1]\tilde{u}_i\tilde{x}$
and $\tilde{I}(\tilde{x})=\tilde{I}_0\theta(\tilde{x})$, where
$\theta(\tilde{x})$ is the unitary step function. In this case all
results are analytical. The general expression of $\tilde{F}$ is
omitted here for brevity, but we give its linear limit obtained for
small excitation probability $\tilde{I}_0$ and velocity $\tilde{V}_B$.
The stalling force is
\begin{equation}
\tilde{F}(0)=-\frac{2}{\tilde{D}_1}\tilde{J}_t+2\left(\frac{1}{\tilde{D}_1}-1\right)
\tilde{\bar{J}}_2,
\label{eqF3}
\end{equation}
with
\begin{eqnarray}
\tilde{J}_t&=&-\frac{\tilde{I}_0\tu_d\tu_1^2
e^{-\tu_1}}{4\Delta^2\left(1-e^{-\tu_1}\right)^2}
\left[\Delta\!+\!\frac{\cosh(\tu_1-\tu_2/2)
-\cosh\delta}{\sinh\delta/(2\delta)}\right]\nonumber\\
\tilde{\bar{J}}_2&=&-\frac{\tilde{I}_0\tu_1\tu_2\tilde{\tau}}{4\Delta}
\left[\frac{1}{\tu_1\tilde{\tau}}
+\frac{\tu_2}{2}-\frac{\cosh\delta\cosh\frac{\tu_1}{2}-\cosh\frac{\tu_d}{2}}
{\sinh\frac{\tu_1}{2}\sinh\delta/\delta}\right]\nonumber
\end{eqnarray}
where $\tu_d=\tu_2-\tu_1$, $\delta=\sqrt{\tu_2^2/4+1/\tilde{\tau}}$, and
$\Delta=\tu_d\tu_1+1/\tilde{\tau}$.

For $U_1=U_2$ and $D_1{\neq}D_2$ (friction fluctuations only, no potential
fluctuations), $\tilde{J}_t=0$ and the stalling force is proportional to
$\tilde{D}_1-1=(D_1-D_2)/D_2$. Therefore, everything else being fixed, the
direction and strength of the force and of the ensuing motion is determined
by the variation of diffusion constant (or friction coefficient) induced by
the state fluctuations.

The nonlinear behaviour of the stalling force $\tilde{F}(0)$ versus
pumping intensity $\tilde{I}_0$ and lifetime $\tilde{\tau}$ is
illustrated in figure 4. The graph shows that there is an optimal
$\tilde{\tau}$, i.e., an optimal ratio between the excited state
lifetime and diffusion time. This optimal value decreases for
increasing $\tilde{I}_0$. This behaviour is similar to that exhibited
by the fluctuating-potential models \cite{prost94}.
\begin{figure}
\begin{center}
\includegraphics{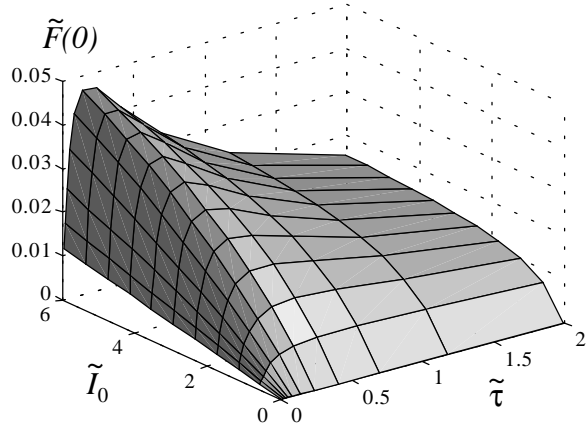}
\end{center}
\caption{Stalling force $\tilde{F}(0)$ versus pumping intensity $\tilde{I}_0$ and
lifetime $\tilde{\tau}$ (dimensionless quantities), for $\tu_1=\tu_2=1$ and
$\tilde{D}_1=D_1/D_2=2$.}
\end{figure}

For nonzero velocity $\tilde{V}_B$, the (linearized) force is
\begin{equation}
\tilde{F}(\tilde{V}_B)=\tilde{F}(0)-\tilde{\eta}_{\mathrm{eff}}\tilde{V}_B,
\end{equation}
where
\begin{equation}
\tilde{\eta}_{\mathrm{eff}}=\frac{1}{\tilde{D}_1}\left[1-\frac{\tu_1^2e^{-\tu_1}}
{(1-e^{-\tu_1})^2}\right].
\end{equation}
Therefore the maximum velocity of the motor is
\begin{equation}
\tilde{V}_{\mathrm{max}}=\frac{\tilde{F}(0)}{(\tilde{\eta}_{\mathrm{eff}}
+\tilde{\eta}_p)}.
\end{equation}

For $U_1{\neq}U_2$ and $D_1=D_2$ (potential fluctuations only, no
friction fluctuations), a nonzero current $\tilde{J}_t$ is
established, i.e., a long-range displacement of body A with respect to
body B occurs. If body B is also fixed (i.e., it is linked to body C),
this case reduces to the ``standard'' one discussed for example in
reference \cite{prost94}. The average speed of body A will then be
$\tilde{v}=2\tilde{J}_t=-\tilde{D}_1\tilde{F}(0)$.

An interesting difference between the two cases of fluctuating
potential and fluctuating friction is given by the behaviour of
$\tilde{F}(0)$ for large potential energies, i.e.,
$\tilde{u}_i=u_i/kT\gg1$. Indeed, when $D_1=D_2$ the force shows a
thermal-activated behaviour, with $\tilde{F}\sim{e^{-\tu_m}}$, where
$\tu_m$ is the minimum of $\tu_1$ and $\tu_2$. On the contrary, when
$D_1{\neq}D_2$ and $u_1=u_2$ a term survives that decays only as
$\tilde{F}\sim1/\tu_1$. This behaviour is evident in figure 5. It can
be explained by considering that for $D_1=D_2$ the force is
proportional to the total average current $\tilde{J}_t$. To contribute
to $\tilde{J}_t$ the motor moving element must overcome the potential
barrier by acquiring sufficient thermal energy and the probability for
this to happen is proportional to $e^{-\tu_i}$. In the case of
$D_1{\neq}D_2$, the force acquires another term proportional to the
average current $\tilde{\bar{J}}_2$ that need not vanish for
$\tilde{J}_t=0$. It corresponds to a circulation of the particle that
moves preferentially forward when in state 1 and backward when in
state 2. However, the motor does not have to reach the potential
maximum and therefore to be thermally activated. This cycling between
the two states and the resulting currents are schematically described
in figure 3.
\begin{figure}
\begin{center}
\includegraphics{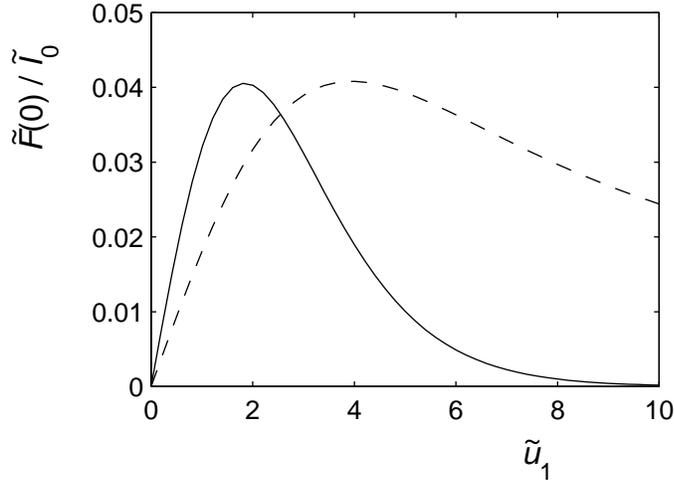}
\end{center}
\caption{Stalling force $\tilde{F}(0)$ (linearized in $\tilde{I}_0$) versus
potential depth $\tu_1$ for the two cases of fluctuating potential
($\tilde{D}_1=D_1/D_2=1$ and $\tu_2/\tu_1=2$, solid line) and fluctuating
friction ($\tu_2/\tu_1=1$ and $\tilde{D}_1=2$, dashed line). In both
examples $\tilde{\tau}=1$.}
\end{figure}

\section{Biological applications: actomyosin motor}
We did not develop this model having in mind a specific biological
application. Nonetheless, it is possible that basic features of some
biological molecular mechanism be captured by our model. The basic working
mechanism of important biological molecular motors is still under debate
\cite{huxley98,huxley98b,kitamura99,yanagida}, so that there is room for
proposing new models and offering different perspectives of this problem.
Therefore we show in the following how our model can be applied to
describing a biological molecular motor. However, we stress that we are not
attempting here to build a ``realistic'' model, our purpose being only to
illustrate the basic idea.

To be specific, let us consider the actin-myosin II system, that is
the molecular motor providing the chemomechanical coupling in muscle
contraction. When a muscle is contracted, thin filaments of actin and
thick filaments of myosin are displaced relative to each other by the
concerted action of many myosin ``heads''. A myosin head is a
molecular unit linked to the myosin filament through a flexible
molecular ``arm'' (or ``neck'') and interacting via weak
intermolecular interactions with the actin filament
\cite{rayment93,rayment93b}. A schematic picture of the actomyosin
system is in figure 6.
\begin{figure}
\begin{center}
\includegraphics{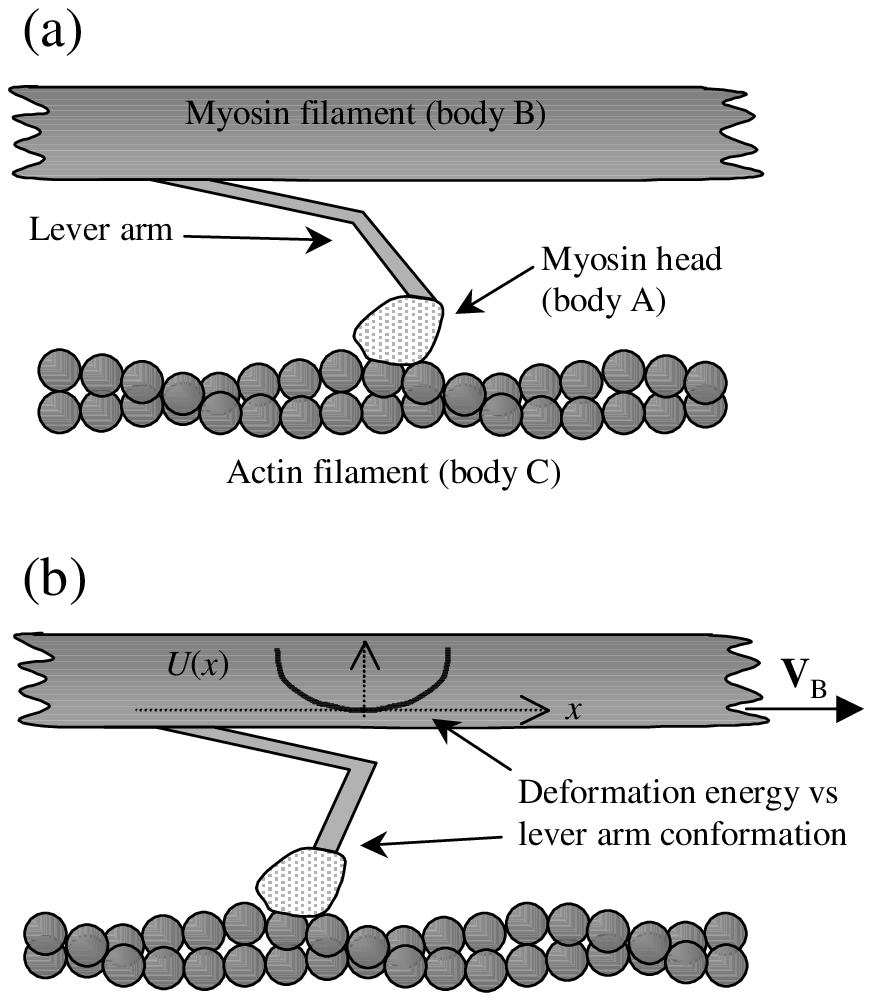}
\end{center}
\caption{Schematic drawing of the actomyosin molecular motor. Drawings (a)
and (b) refer to two different conformations of the lever arm,
corresponding to different values of the coordinate $x$ [$x>0$ in (a),
$x<0$ in (b)]. In (a) are also shown the correspondence rules that link our
model (bodies A,B, and C) to the real system. In (b) is also shown the
single-well energy $U(x)$ characterizing the deformation of the lever arm
versus the deformation coordinate $x$ giving the relative positions of
bodies A and B. The ``Brownian power-stroke'' occurs when the system
switches to the high friction state when $x>0$ (a) and then relaxes its
deformation energy.}
\end{figure}

In reference \cite{prost94} the actomyosin system is modeled by
associating the myosin head with the particle (body A), the actin
filament with the source of potential (body B), and the surrounding
fluid or the actin filament itself with the friction source (body C).
The myosin filament is treated as an additional passive ``load''
dragged on by the moving myosin head. The potential $U(x)$ describing
the adsorption interaction of the head with the actin filament is
periodic and asymmetric reflecting the underlying periodic and
asymmetric structure of the filament. Attaching-detaching of the head
to/from the actin filament are described as sudden changes in the
potential landscape.

Here we propose a different set of correspondence rules in order to
apply our model to the actomyosin system, as shown in figure 6. We
still identify the Brownian particle (body A) with the myosin head.
However, we model here the weak interaction between A and the actin
filament as molecular friction, thus identifying the actin filament
with body C. In a limit case, the low friction state allows almost
free sliding of A on C, while the high friction state impedes any
sliding. The myosin filament plays instead the role of body B. The
potential $U(x)$ is used here to model the deformation energy of the
molecular arm connecting the myosin filament with the head. The
function $U(x)$ therefore needs not be periodic in this case, as the
limited range of arm deformation constrains the particle A into a
single potential well. The periodic potential formalism can be still
used, however, by setting infinitely high energy barriers at ${\pm}L$.

Adenosinetriphosphate (ATP) binding on the myosin head and its
subsequent hydrolysis triggers the variation of molecular friction
coefficient of the myosin head (from high to low friction). By means
of some suitable mechanism (specific intermolecular interactions),
only when the conformation of the myosin arm is bent in a given
direction, for example negative $x$ (figure 6b), the molecule
enzymatic activity is large and ATP binding and hydrolysis may occur.
This corresponds roughly to having $I(x)>0$ for $x<0$ (figure 6b) and
$I(x)=0$ for $x>0$ (figure 6a), where the rate $I$ is proportional to
ATP concentration. After an average time $\tau$, the hydrolysis
reaction has gone through its full cycle and the motor switches back
to a high friction state.

Given these correspondence rules, our model predicts a continuous
directional motion of the myosin motor (head plus filament, i.e.,
bodies A and B) with respect to the actin filament (body C) (of course
this is a relative motion: actually none of the three bodies can be
considered fixed). The general behaviour of the motor as a function of
external load and ATP concentration is in qualitative agreement with
experimental data \cite{spudich94,howard97}.

We emphasize that no direct causal link between the ATP hydrolysis and
conformational transformations is introduced in our model, i.e., the
conformational changes, described in our model by the continuous
variable $x$, are not chemically-driven but are governed solely by
Brownian motion. The asymmetric transition rates allow the elastic
energy stored during the Brownian motion to be converted into directed
long-range motion. In particular the main ``force-generating step''
occurs when the system happens to switch back to the high friction
state for $x>0$ (figure 6a), relaxing then its deformation energy $U$
by moving back to $x\approx0$. This step may be considered a sort of
``Brownian power-stroke''. In contrast, our model assumes no
``chemically-driven power-stroke''. In this sense, it differs from
many other models of biological motors. A power-stroke is often
modeled either as a sudden change of the elastic constant
characterizing the potential \cite{mogilner98}, occurring as a
consequence of ATP-induced state transitions, or as free sliding on an
asymmetric potential as in Brownian ratchet models \cite{prost94}. On
the other hand, the force-generation mechanism of our model in the
limit of very high friction in state 2 and very low friction in state
1 becomes similar to that of the old model by A. F. Huxley
\cite{huxley57}. A similar sort of ``Brownian power-stroke'' is
assumed also in references \cite{oester95,oester95b}, in connection
with the modeling of kinesin.

\section{Conclusions}
The model presented in this paper was actually developed for a
specific application, namely the recently discovered strong
enhancement of the light-induced molecular reorientation taking place
in dye-doped nematic liquid crystals \cite{janossy90}. A more detailed
treatment of our modeling of this phenomenon has been already
published elsewhere \cite{kreuzer00}, and here we limit ourselves to a
brief discussion. The relationship between this optical effect and
molecular motors was first pointed out by Palffy-Muhoray and Weinan,
without introducing however the possibility of an active role of
fluctuating friction \cite{palffy98}. In our model, dye molecules play
the role of motor particles (body A) and the nematic molecular
director plays the role of potential source (body B). The coordinate
$x$ is here the angle between the dye molecule orientation and the
molecular director, so that the motor is a rotary one. The role of the
friction source C is played by the translational (centre-of-mass)
degrees of freedom of liquid crystal molecules themselves. Light
powers the motor by continuously promoting internal electronic
transitions in the dye molecules, and the final output is a torque
(corresponding to the force $F$) acting on the molecular director and
eventually causing its reorientation. There is substantial
experimental evidence \cite{marrucci97b,marrucci98} that the
rotational friction coefficients of dye molecules in the two
electronic states involved in the transitions are quite different, so
that the fluctuating-friction mechanism is indeed at work. This proves
that this hypothetical mechanism is actually realistic, at least in
one specific example.

In this paper, our main goal was to illustrate in a general abstract form,
independent of any specific application, the idea that a stochastic
fluctuation of a kinetic coefficient, such as friction, mobility, or
diffusion constant, can be an effective mechanism for converting chemical
(or light) energy into directional motion and mechanical work at the
molecular scale, at which Brownian fluctuations dominate. The possibility
of extracting work from Brownian motion by means of a suitable modulation
of kinetic properties, as opposed to equilibrium potential forces, appears
to be new, at least within the field of molecular motors.

We believe that this concept can be profitable in several other
fields, besides the nonlinear optics one. In particular, as discussed
in section IV, many biological motors are still waiting for a detailed
modeling, and any new idea can be very useful in the quest for
complete understanding of these complex systems. Similarly, in the
field of nanotechnology, the possibility of driving transport via a
suitable modulation of the microscopic kinetic coefficients can be an
interesting design concept.

In all cases, the relevance of this idea is further enhanced if one takes
into account the high structural sensitivity of kinetic coefficients
occurring commonly in activated systems, where small changes in the
activation energy can lead to huge variations of the kinetic coefficients.

We close by adding that the fluctuating-friction motor idea is not limited
to stochastic fluctuations. Just as in the case of fluctuating-potential
motors, a periodic modulation in time of the friction coefficient is also
expected to induce similar phenomena. The modulation period will then have
to be close to the typical diffusion time of the particle, in order to be
effective, leading to a new kind of stochastic resonance phenomenon.

\ack
This work was realized within the framework of the European
Brite-Euram Network LC-PHOTONET.

\end{document}